\documentclass{article}

\usepackage{fullpage}
\usepackage{amsmath}
\usepackage{bm}
\usepackage{graphicx}
\usepackage{caption}
\usepackage{subcaption}
\usepackage{hyperref}

\let\oldbibliography\thebibliography
\renewcommand{\thebibliography}[1]{%
  \oldbibliography{#1}%
  \setlength{\itemsep}{0pt}%
}

\title{Mathematics is Physics}

\author{M. S. Leifer \\ Perimeter Institute for Theoretical Physics \\
31 Caroline St. N., Waterloo, ON, Canada N2L 2Y5}

\date{February 19, 2015}

\begin{document}

\maketitle

\begin{abstract}
  In this essay, I argue that mathematics is a natural science---just
  like physics, chemistry, or biology---and that this can explain the
  alleged ``unreasonable'' effectiveness of mathematics in the
  physical sciences.  The main challenge for this view is to explain
  how mathematical theories can become increasingly abstract and
  develop their own internal structure, whilst still maintaining an
  appropriate empirical tether that can explain their later use in
  physics.  In order to address this, I offer a theory of mathematical
  theory-building based on the idea that human knowledge has the
  structure of a scale-free network and that abstract mathematical
  theories arise from a repeated process of replacing strong analogies
  with new hubs in this network.  This allows mathematics to be seen
  as the study of regularities, within regularities, within \ldots,
  within regularities of the natural world.  Since mathematical
  theories are derived from the natural world, albeit at a much higher
  level of abstraction than most other scientific theories, it should
  come as no surprise that they so often show up in physics.

  This version of the essay contains an addendum responding to Slyvia
  Wenmackers' essay \cite{Wenmackers2015} and comments that were made
  on the FQXi website \cite{MP2015}.
\end{abstract}

\section{The Unreasonable Effectiveness of Mathematics in the Physical
Sciences}

\begin{quotation}
  The miracle of the appropriateness of the language of mathematics
  for the formulation of the laws of physics is a wonderful gift which
  we neither understand nor deserve.  --- Eugene Wigner
  \cite{Wigner1960}
\end{quotation}

Mathematics is the language of physics, and not just any mathematics
at that.  Our fundamental laws of physics are formulated in terms of
some of the most advanced and abstract branches of mathematics.
Seemingly abstruse ideas like differential geometry, fibre bundles,
and group representations have been commonplace in physics for
decades, and theoretical physics is only getting more abstract, with
fields like category theory playing an increasingly important role in
our theory building.  Do we simply have to accept this mathematization
as a brute fact about our universe, or can it be explained?

My thesis is that mathematics is a natural science---just like
physics, chemistry, or biology---albeit one that is separated from
empirical data by several layers of abstraction.  It is nevertheless
fundamentally a theory about our physical universe and, as such, it
should come as no surprise that our fundamental theories of the
universe are formulated in terms of mathematics.

The philosophical worldview underlying my arguments is that of
naturalism \cite{Quine1969, Papineau2009}.  Naturalism is the position
that everything arises from natural properties and causes, i.e.\
supernatural or spiritual explanations are excluded.  In particular,
natural science is our best guide to what exists, so natural science
should guide our theorizing about the nature of mathematical objects.

Mathematics is not usually thought of as an empirical science, so
naturalism may seem irrelevant to its philosophy.  However, we have
one rather conspicuous empirical data point about it, namely the
alleged ``unreasonable'' effectiveness of mathematics in the physical
sciences.  Since our fundamental laws of physics are formulated in
terms of some of the most advanced branches of mathematics, a
philosophy of mathematics that explains this should be preferred to
one in which it is an ``unreasonable'' accident or miracle.  Such a
theory would also be falsifiable in the sense that, were it to be the
case our future fundamental theories of physics resist
mathematization, being only explainable in words or only formalizable
in terms of very elementary mathematics, then our philosophy of
mathematics would have been proved wrong.

The philosophy of mathematics I develop here is closely related to
those of Quine and Putnam \cite{Quine1951, Putnam1967, Putnam1968,
  Putnam1972}, who instigated the naturalistic approach to mathematics
and even suggested that logic could be empirical.  However, unlike
them, I am inclined towards a more pragmatic theory of truth so, for
me, abstract mathematical objects can be called real insofar as they
are useful for our scientific reasoning.  This evades the problem of
trying to find direct referrents of mathematical objects in the
physical world.  Instead, I simply have to explain why they are
useful.  In order to do this, I shall have to investigate where
mathematics comes from.  In this vein, I shall argue that human
knowledge has the structure of a \emph{scale-free network} and offer a
theory of mathematical theory-building that emphasizes reasoning by
analogy within this network.  This explains how mathematics can become
increasingly abstract, whilst maintaining its tether to empirical
reality.  It also explains why abstract areas of mathematics that are
developed in seeming isolation from physics often show up later in our
physical theories.

\begin{figure}[htb]
  \centering
  \begin{subfigure}[t]{0.45\textwidth}
    \includegraphics[width=\textwidth]{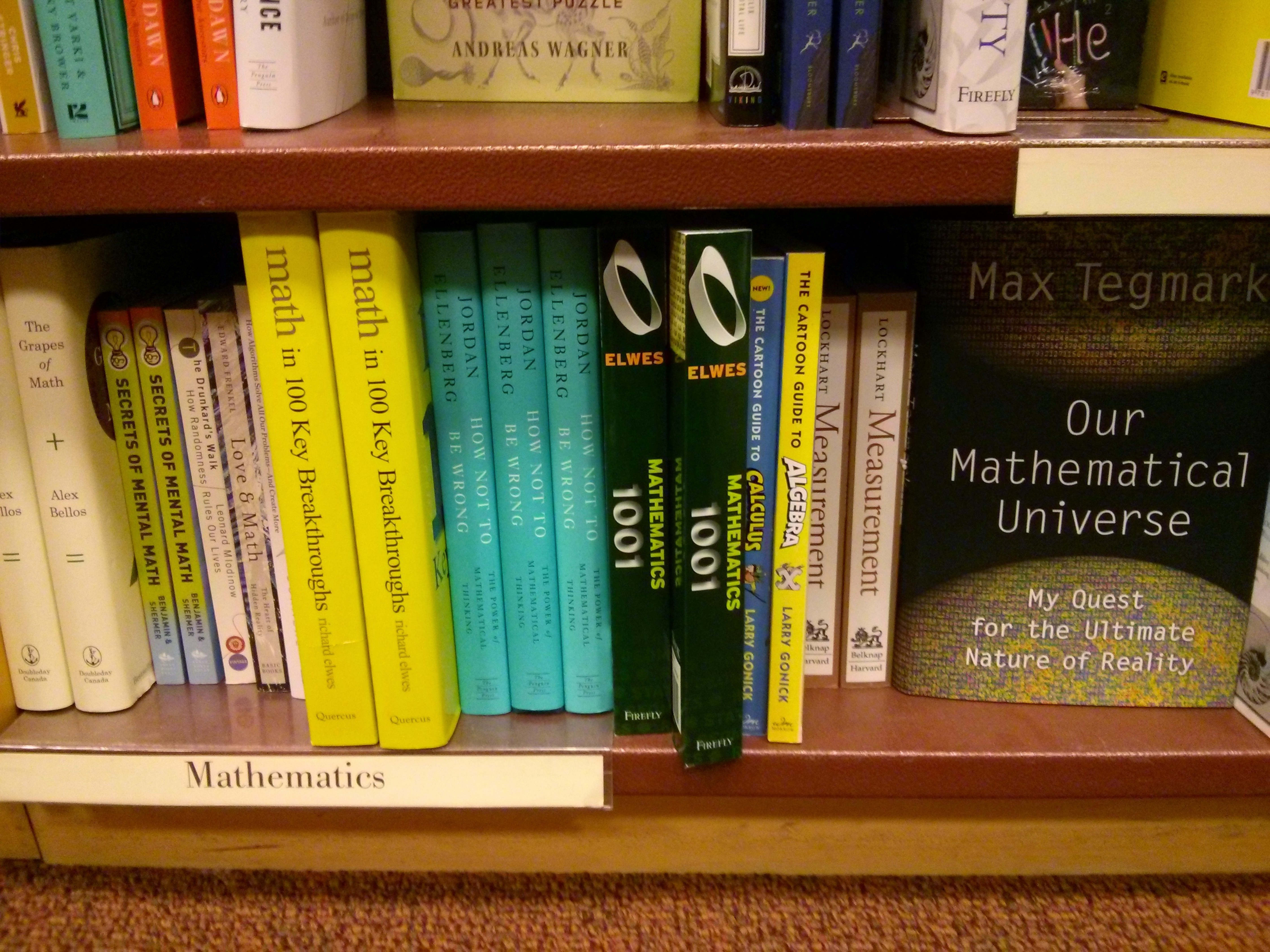}
    \caption{The staff at my local bookstore agree with Tegmark's
    Mathematical Universe Hypothesis, as they have displayed his book
    about the nature of our physical universe in the mathematics
    section.}
  \end{subfigure}
  \qquad
  \begin{subfigure}[t]{0.45\textwidth}
    \includegraphics[width=\textwidth]{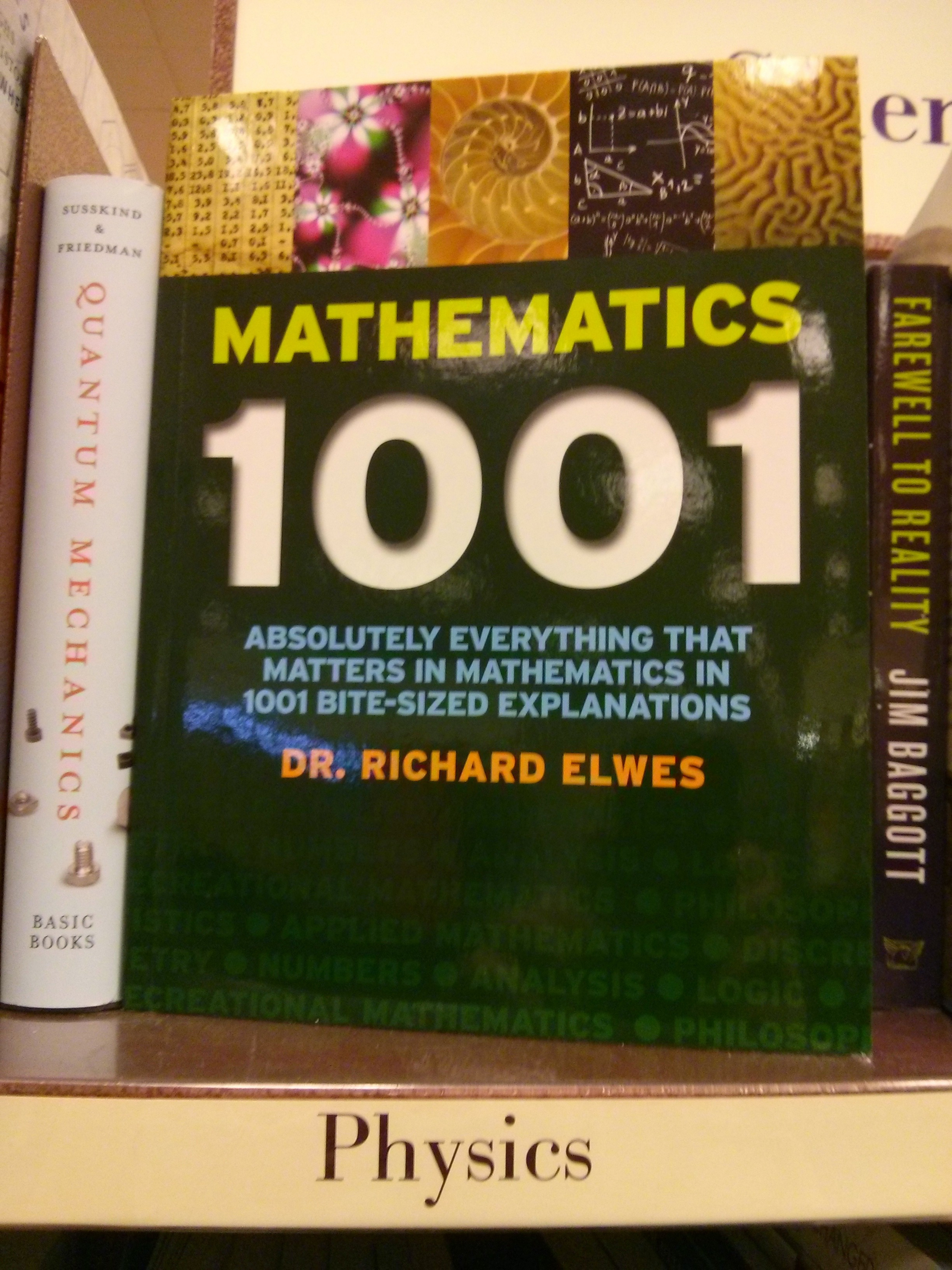}
    \caption{In response, I placed a mathematics book in the physics
      section, in order to illustrate my counter-hypothesis.}
  \end{subfigure}
  \caption{\label{compare}Contrasting the Mathematical Universe
    Hypothesis with my view that mathematics is derived from the
    physical universe.}
\end{figure}

My title, ``Mathematics is Physics'', is deliberately chosen in
contrast to our dear leader Max Tegmark's Mathematical Universe
Hypothesis \cite{Tegmark2014} (see Figure~\ref{compare} for a
comparison of the two views).  This asserts that our universe is
nothing but a mathematical structure and that all possible
mathematical structures exist in the same sense as our universe.  The
first part of the hypothesis may be condensed to ``Physics is
Mathematics'', so in this sense I am arguing for precisely the
opposite.  I will compare and contrast the two ideas at the end.

\section{What is mathematics?}

\label{math}

The philosophy of mathematics is now a vast and sprawling
subject\footnote{See \cite{Horsten2015} for an overview of the
  subject}, so I cannot possibly cover all views of the nature of
mathematics here.  However, it is useful to introduce two of the more
traditional schools of thought---mathematical platonism and
formalism---to serve as foils to the naturalistic theory I then
develop.

\subsection{Mathematical platonism}

Mathematical platonism is the idea that mathematical objects are
abstract entities that exist objectively and independently of our
minds and the physical universe.  For example, the geometric concept
of a straight line does not refer to any approximation of a straight
line that we might draw with pencil on paper.  If we view any real
world approximation through a microscope we will see that it has rough
edges and a finite thickness.  On the other hand, a geometric straight
line is perfectly straight and has no thickness.  It is a
fundamentally one-dimensional object, which we can only ever
approximate in our three-dimensional universe.  Mathematical objects
like straight lines do not exist in our universe, so the platonist
asserts that they exist in an abstract realm, somewhat akin to the
Platonic world of forms.  They further assert that we have direct
access to this realm via our mathematical intuition.

Mathematical platonism is in direct conflict with naturalism.  A
naturalistic theory has no place for a dualistic mind that is
independent of the structure of our brains.  Therefore, if we have
intuitive access to an abstract realm, our physical brains must
interact with it in some way.  Our best scientific theories contain no
such interaction\footnote{This argument is known as Benacerraf's
  epistemological objection to mathematical platonism
  \cite{Benacerraf1983}.}.  The only external reality that our brains
interact with is \emph{physical} reality, via our sensory experience.
Therefore, unless the platonist can give us an account of where the
abstract mathematical realm actually is in physical reality, and how
our brains interact with it, platonism falls afoul of naturalism.

Furthermore, our brains are the product of evolution by natural
selection, so, on the naturalist view, whatever mathematical
intuitions we have are either a reflection of what was useful for
survival, or products of the general intelligence that evolution has
endowed us with.  Our evolutionary instincts are often poor guides to
reality, so it is difficult to see how mathematics could be objective
if mathematical intuition is of this type.  If it is instead a product
of general intelligence then mathematics could either be akin to a
creative work of fiction, or it must be the result of reasoning about
the physical world that we find ourselves in.  Only the latter can
explain why our theories of physics are mathematical, so this account
should be preferred.

\subsection{Formalism}

Formalism is the view that mathematics is just a game about the formal
manipulation of symbols.  We specify some symbols, such as marks on a
sheet of paper, and rules for deriving one string of symbols from
another.  Anything that can be derived from those rules is a theorem
of the resulting mathematical system.

As a toy example, we could posit that the symbols are $0$ and $1$.
The rules are that you may replace the empty string with either $0$ or
$1$, you may replace any string $s$ ending in $0$ with $s1$, and any
string $s$ ending in $1$ with $s0$.  $10 \rightarrow 1010101$ is an
example of a (very boring) theorem in this (very boring) formal
system.

Formalism has the advantage that it untethers mathematics from an
abstract objective realm that is independent of our minds.  It is
therefore naturalistic in the sense that it does not posit a
supernatural world.  Mathematics instead becomes an intellectual game,
in which we may posit any rules we like.  However, formalism has two
difficult obstacles to deal with.  Firstly, mathematicians do not
study arbitrary formal systems.  There is no ``adding zeroes and ones
to the end of binary strings'' research group in any mathematics
department.  Formalists must specify which formal systems count as
mathematics.  Why is group theory a branch of mathematics, but adding
zeroes and ones to the end of binary strings is not?  Secondly, even
if this is achieved, the formalist has no explanation of why abstract
branches of mathematics show up in physics.

\subsection{Mathematics as a natural science}

In response to the objections to platonism and formalism, I wish to
defend the idea that mathematics is a natural science, i.e.\ its
subject matter is ultimately our physical universe.  To do this, I
will borrow one idea each from platonism and formalism.  Along with
the platonists, I want to view mathematics as being about an objective
world that exists independently of us.  The difference is that, in my
case, this is just the actual physical world that we live in, rather
than some hypothetical abstract world of forms.

However, mathematical objects are more abstract than those that appear
in the physical world, and they include entities that seem to have no
physical referrent, such as hierarchies of infinities of ever
increasing size.  To deal with this issue, I maintain that
mathematical objects do not refer directly to things that exist in the
physical universe.  As the formalists suggest, mathematical theories
are just abstract formal systems, but not all formal systems are
mathematics.  Instead, mathematical theories are those formal systems
that maintain a tether to empirical reality through a process of
abstraction and generalization from more empirically grounded
theories, aimed at achieving a pragmatically useful representation of
regularities that exist in nature.

It is relatively easy to defend the idea that elementary concepts like
the finite natural numbers are theories of things that actually exist
in the world, viz.\ the rules of arithmetic are derived from what
actually happens if you combine collections of discrete objects such
as sheep, rocks, or apples.  However, the advanced theories of
mathematics deal with entities that are not related to the physical
world in any obvious way at all.  For example, nobody has ever seen an
infinite collection of sheep of any type, let alone one that has the
cardinality of some specific transfinite number.  Moreover,
mathematical theories seem to have their own autonomy, independent of
the natural sciences.  Pure mathematicians have their own programs of
research, with well-motivated questions that are internal to their
mathematical theories, making no obvious reference to empirical
reality.  In response to this, some naturalists are content to only
deal with the mathematics that actually occurs in science, and it has
been suggested that the theory of functions of real variables is
sufficient for this.  However, much more abstract mathematics than
this appears in modern physics and, in any case, if we are to explain
why the fruits of pure mathematics research so often appear in physics
at a later date, then we are going to need a theory that encompasses
purely abstract research.  So, for me, the biggest problem is to
explain how abstract pure mathematical theories can be empirical
theories in disguise, and to do so in such a way that the later
application of these theories to physics becomes natural.

This question cannot be answered without looking at where mathematical
theories come from.  If I can argue that mathematical theories
maintain an appropriate tether to empirical reality then it should be
no surprise that the regularities encoded within them, which already
refer to nature, later show up in natural science.  Developing a
theory of human knowledge and mathematical theory-building that does
this is the main challenge for my approach.

\section{The structure of human knowledge}

Although we are concerned with how mathematics relates to the physical
world, it is important to realize that all our knowledge is
\emph{human} knowledge, i.e.\ it is discovered, organized, learned,
and evaluated in an ongoing process by a social network of finite
beings.  This means that the structure of our knowledge will, in
addition to reflecting the physical world, also reflect the nature of
the process that generates it.  By this I do not mean to imply that
human knowledge is merely a social construct---it is still knowledge
about an objective physical world.  However, if we are to explain why
abstract pure mathematics later shows up in physics, we are going to
have to examine the motivations and methodology of those who develop
that mathematics.  We have to uncover a hidden empirical tether in
their methods and explain how it can be that the patterns and
regularities they are studying are in fact patterns and regularities
of nature in disguise.

\begin{figure}[htb]
  \centering
  \includegraphics[scale=0.5]{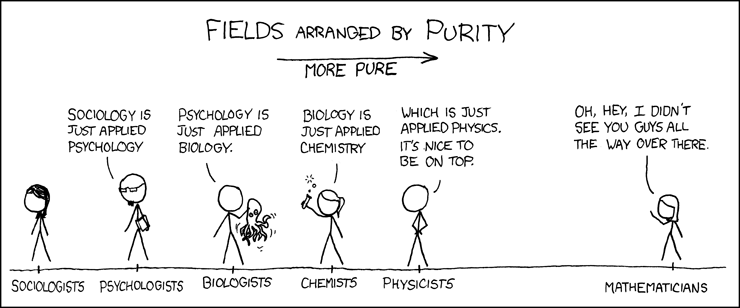}
  \caption{\label{xkcd}This xkcd cartoon depicts how many people view
    the structure of knowledge, organized hierarchically from more
    fundamental to more applied\cite{Munroe2008}.}
\end{figure}

It is common to view the structure of human knowledge as hierarchical,
as satirized by the xkcd cartoon in Figure~\ref{xkcd}.  The various
attempts to reduce all of mathematics to logic or arithmetic reflect a
desire view mathematical knowledge as hanging hierarchically from a
common foundation.  However, the fact that mathematics now has
multiple competing foundations, in terms of logic, set theory or
category theory, indicates that something is wrong with this view.

\begin{figure}[htb]
  \centering
  \includegraphics[scale=0.5]{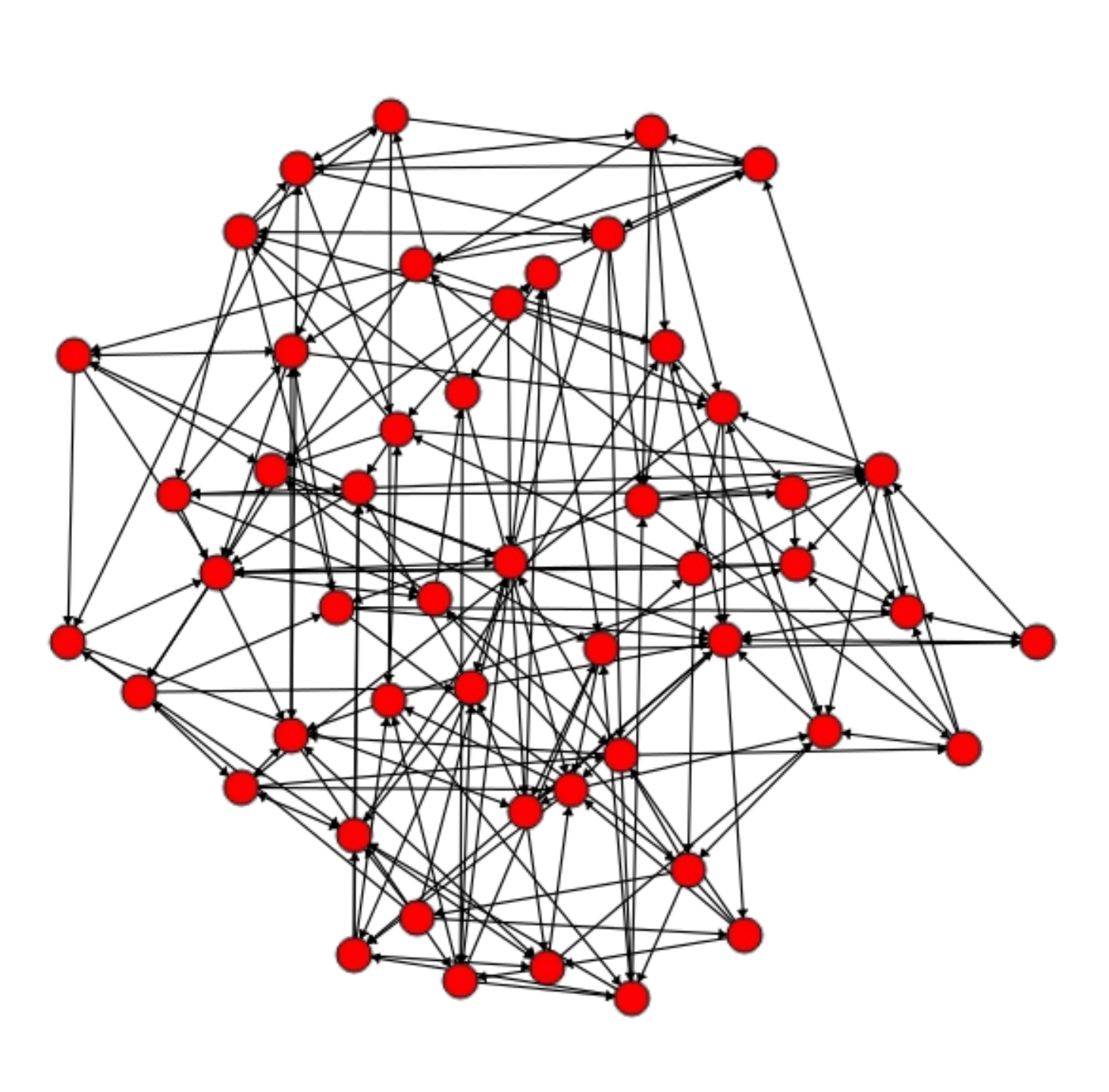}
  \caption{\label{network}An example of a network with nodes (red
    circles) and links between them (arrows).  This network was
    generated by the SocNetV software package \cite{socnetv}.}
\end{figure}

Instead of a hierarchy, we are going to attempt to characterize the
structure of human knowledge in terms of a network consisting of nodes
with links between them (see Figure~\ref{network}).  Roughly speaking,
the nodes are supposed to represent different fields of study.  This
could be done at various levels of detail.  For example, we could draw
a network wherein nodes represent things like ``physics'' and
``mathematics'', or we could add more specific nodes representing
things like ``quantum computing'' and ``algebraic topology''.  We
could even go down to the level having nodes representing individual
concepts, ideas, and equations.  I do not want to be too precise about
where to set the threshold for a least digestible unit of knowledge,
but to avoid triviality it should be set closer to the level of
individual concepts than vast fields of study.

Next, a link should be drawn between two nodes if there is a strong
connection between the things they represent.  Again, I do not want to
be too precise about what this connection should be, but examples
would include an idea being part of a wider theory, that one thing can
be derived from the other, or that there exists a strong direct
analogy between the two nodes.  Essentially, if it has occurred to a
human being that the two things are strongly related, e.g.\ if it has
been thought interesting enough to do something like publish an
academic paper on the connection, and the connection has not yet been
explained in terms of some intermediary theory, then there should be a
link between the corresponding nodes in the network.

If we imagine drawing this network for all of human knowledge then it
is plausible that it would have the structure of a \emph{scale-free
  network}\cite{Albert2002}.  Without going into technical details,
scale-free networks have a small number of \emph{hubs}, which are
nodes that are linked to a much larger number of nodes than the
average.  This is a bit like the $1\%$ of billionaires who are much
richer than the rest of the human population.  If the knowledge
network is scale-free then this would explain why it seems so
plausible that knowledge is hierarchical.  In a university degree one
typically learns a great deal about one of the hubs, e.g.\ the hub
representing fundamental physics, and a little about some of the more
specialized subjects that hang from it.  As we get ever more
specialized, we typically move away from our starting hub towards more
obscure nodes, which are nonetheless still much closer to the starting
hub than to any other hub.  The local part of the network that we know
about looks much like a hierarchy, and so it is not surprising that
physicists end up thinking that everything boils down to physics
whereas sociologists end up thinking that everything is a social
construct.  In reality, neither of these views is right because the
global structure of the network is not a hierarchy.

As a naturalist, I should provide empirical evidence that human
knowledge is indeed structured as a scale-free network.  The best
evidence that I can offer is that the structure of pages and links on
the World Wide Web and the network of citations to academic papers are
both scale free \cite{Albert2002}.  These are, at best, approximations
of the true knowledge network.  The web includes facts about the
Kardashian family that I do not want to categorize as knowledge, and
not all links on a website indicate a strong connection, e.g.\
advertising links.  Similarly, there are many reasons why people cite
papers other than a direct dependence.  However, I think that these
examples provide evidence that the information structures generated by
a social network of finite beings are typically scale-free networks,
and the knowledge network is an example of such a structure.

\section{A theory of theory-building}

We are now at the stage where I can explain where I think mathematical
theories come from.  The main idea is that when a sufficiently large
number of strong analogies are discovered between existing nodes in
the knowledge network, it makes sense to develop a formal theory of
their common structure, and replace the direct connections with a new
hub, which encodes the same knowledge more efficiently.

\begin{figure}[htb]
  \begin{subfigure}[t]{0.45\textwidth}
    \includegraphics[width=\textwidth]{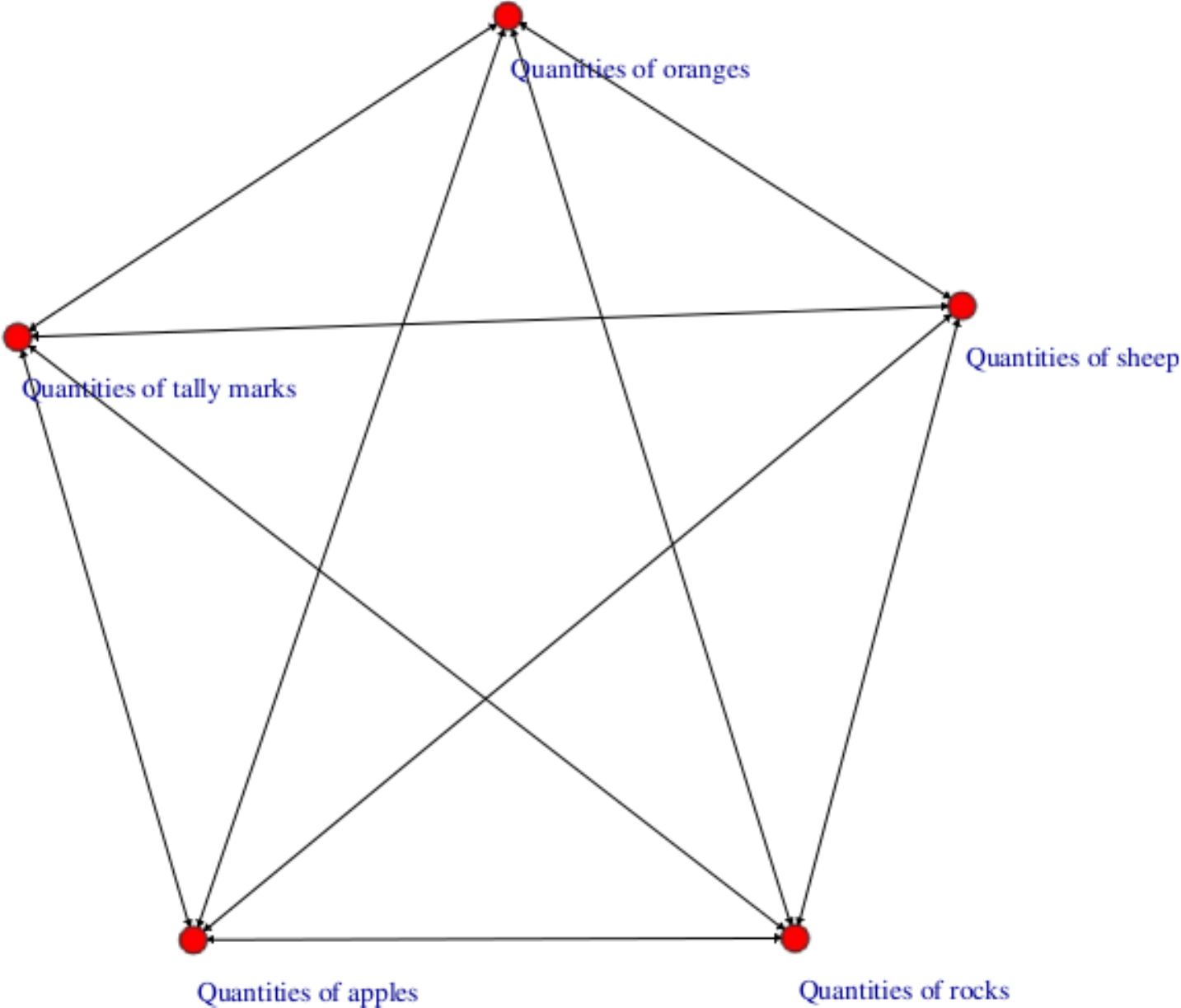}
    \caption{Absent a theory of number, there are a large number of
      direct analogies between quantities of discrete objects like
      sheep, rocks, apples, etc.  Only analogies between five types of
      discrete objects are depicted here.  There would be vastly more
      in the real knowledge network.}
  \end{subfigure}
  \qquad
  \begin{subfigure}[t]{0.45\textwidth}
    \includegraphics[width=\textwidth]{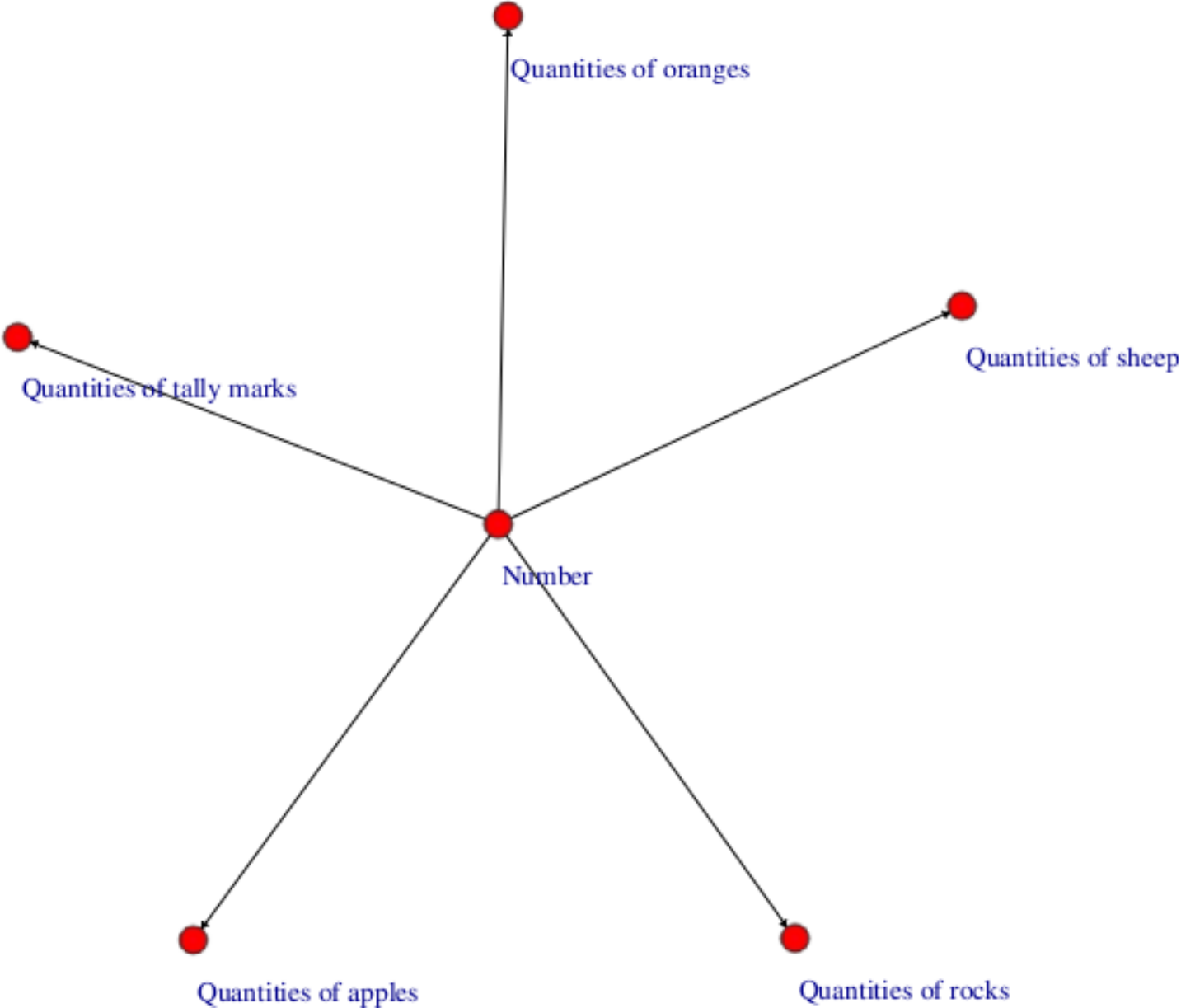}
    \caption{A new hub is introduced to capture the common structure
      indicated by the analogies. The theory of number captures the
      regularity at one higher level of abstraction.}
  \end{subfigure}
  \caption{\label{analogy}How abstract mathematical theories arise
    from analogies between theories at a lower level of abstraction,
    which are closer to the empirical data.}
\end{figure}

As a first example, consider the following just-so story about where
natural numbers and arithmetic might have come from.  Initially,
people noticed that discrete quantities of sheep, rocks, apples, etc.\
all have a lot of properties in common.  Absent a theory of this
common structure, the network of knowledge has a vast number of direct
connections between the corresponding nodes (see
Figure~\ref{analogy}).  It therefore makes sense to introduce a more
abstract theory that captures the common features of all these things,
and this is where the theory of number comes in.  A vast array of
individual connections is replaced by a new hub, which has the effect
of organizing knowledge more efficiently.  Now, instead of having to
learn about quantities of sheep, rocks, apples, etc.\ individually and
then painstakingly investigate each analogy, one need only learn about
the theory of number and then apply it to each individual case as
needed.  In this way, the theory of number remains essentially
empirical.  It is about regularities that exist in nature, but is
removed from the empirical data by one layer of abstraction compared
to our direct observations.

Once it is established, the theory of number allows for the
introduction of new concepts that are not present in finite
collections of sheep, such as infinite sequences and limits.  As the
theory is more abstract than the empirical phenomena it is derived
from, it develops its own internal life and is partially freed from
its empirical ties.  Such internal questions are sometimes settled by
pragmatic considerations, e.g.\ which definitions make for the most
usable extension of the theory beyond strictly finite quantities, but
mostly by the way in which the theory ends up interacting with the
larger structure of mathematical knowledge.  For example, the question
of how to define limits of infinite sequences was not really settled
until those definitions were needed to understand calculus and
analysis, which are themselves abstractions of the physically rooted
geometries and flows that exist in our universe.

Once several abstract theories have been developed, the process can
continue at a higher level.  For example, category theory was born out
of the strong analogies that exist between the structure preserving
maps in group theory, algebraic topology, and homological algebra
\cite{Landry2005}.  At first sight, it seems like this is a
development that is completely internal to pure mathematics, but
really what is going on is that mathematicians are noticing
regularities, within regularities, within regularities \ldots, within
regularities of the physical world.  In this way, mathematics can
become increasingly abstract, and develop its own independent
structure, whilst maintaining a tether to the empirical world.

Now, to be sure, in the course of this process mathematicians have to
make several pragmatic choices that seem to be independent of physics.
For example, is it better to reject the axiom of choice, which says
that when you have several sets of objects (including a possibly
infinite number of sets) then there is a way of picking one object
from each of them, or is it better to accept the counter-intuitive
implication of this that if you peel a mathematically idealized orange
then the peel can be used to completely cover two mathematically
idealized oranges of the same size\footnote{This is the Banach--Tarski
  paradox \cite{Banach1924}.}?  Ultimately, the axiom of choice is
accepted by most mathematicians because it leads to the most useful
theories.  The theories that employ it lead to a more efficient
knowledge network than those that reject it, and this trumps any
apparent physical implausibility.  In my view, intuition for efficient
knowledge structure, rather than intuition about an abstract
mathematical world, is what mathematical intuition is about.

It is now straightforward to explain why abstract mathematical
theories show up so often in physics.  Abstract mathematical theories
are about regularities, within regularities, within \ldots, within
regularities of our physical world.  Physical theories are about
exactly the same thing.  The only difference is that whilst
mathematics started from empirical facts that only required informal
observations, physics includes the much more accurate empirical
investigations that only became possible due to scientific and
technological advances, e.g.\ the development of telescopes and
particle colliders.  Nonetheless, it should be no surprise that
regularities that are about the selfsame physical world turn out to be
related.  It should also be no surprise that pure mathematicians often
develop the mathematics relevant for physics a long time before it is
needed by physicists, as they have had much more time to investigate
regularities at a higher level of abstraction.

\section{Implications}

Before concluding, I want to describe two implications of my theory of
knowledge and mathematics for the future of physics.

Firstly, in network language, the concept of a ``theory of
everything'' corresponds to a network with one enormous hub, from
which all other human knowledge hangs via links that mean ``can be
derived from''.  This represents a hierarchical view of knowledge,
which seems unlikely to be true if the structure of human knowledge is
generated by a social process.  It is not impossible for a scale-free
network to have a hierarchical structure like a branching tree, but it
seems unlikely that the process of knowledge growth would lead
uniquely to such a structure.  It seems more likely that we will
always have several competing large hubs and that some aspects of
human experience, such as consciousness and why we experience a unique
present moment of time, will be forever outside the scope of physics.

Nonetheless, my theory suggests that the project of finding higher
level connections that encompass more of human knowledge is still a
fruitful one.  It prevents our network from having an unwieldy number
of direct links, allows us to share more common vocabulary between
fields, and allows an individual to understand more of the world with
fewer theories.  Thus, the search for a theory of everything is not
fruitless; I just do not expect it to ever terminate.

Secondly, my theory predicts that the mathematical representation of
fundamental physical theories will continue to become increasingly
abstract.  The more phenomena we try to encompass in our fundamental
theories, the further the resulting hubs will be from the nodes
representing our direct sensory experience.  Thus, we should not
expect future theories of physics to become less mathematical, as they
are generated by the same process of generalization and abstraction as
mathematics itself.

\section{Conclusions}

I have argued that viewing mathematics as a natural science is the
only reasonable way of understanding why mathematics plays such a
central role in physics.  I have also offered a theory of mathematical
theory-building that can explain how mathematical theories maintain a
tether to the physical world, despite becoming ever more abstract.

To conclude, I want to contrast my theory with Tegmark's Mathematical
Universe Hypothesis.  In my theory, abstract mathematics connects to
the physical world via our direct empirical observations.  The latter
are at the very edges of the knowledge network, as far away from our
most abstract mathematical theories as they possibly could be.  They
are the raw material from which our mathematical theories are
constructed, but the theories themselves are just convenient
representations of the regularities, within regularities, within
\ldots, within regularities of the physical world.  Thus, my view is
opposite to Tegmark's.  Mathematics is constructed out of the physical
world rather than the other way round.

Like my proposal, the Mathematical Universe Hypothesis is
naturalistic.  It asserts that there is no abstract mathematical realm
independent of our physical universe, because the physical world is
\emph{identified} with the realm of all possible mathematical
structures.  However, the universe we find ourselves in is just one
structure in a multiverse of equally real possibilities, and Tegmark's
theory does not explain how we could come to know about these other
mathematical structures.  Indeed, the universes within a multiverse
should not have a strong interaction with each other, otherwise they
would not be identifiable as independent universes, so it is difficult
to see how there could be any causal connection to explain our
abstract mathematical theorizing.

The main evidence for the Mathematical Universe Hypothesis is that
physics is becoming ever more abstract and mathematical, so it looks
like the world described by physics can be identified with a
mathematical structure.  However, my theory provides an explanation
for the increasing abstraction of physics, and can also account for
mathematical theory-building in a natural way so, at present, I think
it should be preferred.

\appendix

\section{Addendum}

I have been asked to update this essay in light of the discussion that
occurred on the FQXi website during the competition.  I cannot
possibly cover all of the comments, so I shall focus on a few that I
consider interesting and accessible.  For more details, it is worth
reading the comment thread in its entirety \cite{MP2015}.

Since Sylvia Wenmackers' winning essay also argues for a naturalistic
approach to mathematics \cite{Wenmackers2015}, I also thought it
worthwhile to discuss how my position differs from hers.  We start
with that before moving on to the website comments.

\subsection{Response to Sylvia Wenmackers' essay}

The main advantage of a naturalistic view of mathematics is that it
offers a simple explanation of the alleged unreasonable effectiveness
of mathematics in physics.  If mathematics is fundamentally about the
physical world then it is no surprise that it occurs in our
description of the physical world.

However, the naturalistic view is prima facie absurd because
mathematical truth seems to have nothing to do with physical world.
The main task of a naturalistic account of mathematics is therefore a
deflationary one: explain why mathematical theories are empirical
theories in disguise and why we have been so easily misled into
thinking this is not the case.  Wenmackers gives a battery of
arguments for this position based on four ``elements''.  Here, I will
focus on the first and third elements, which are based on evolution by
natural selection and selection bias respectively, because this is
where I disagree with her.  I more or less agree with her discussion
of the other two elements.

\subsubsection{The evolutionary argument}

The evolutionary argument asserts that our cognitive abilities, and in
particular our ability to do mathematics, are the result of evolution
by natural selection.  It is therefore no surprise that they reflect
physical reality, as physical reality provided the environmental pressures
that selected for those abilities.

This argument works well for basic mathematics, such as arithmetic and
elementary pattern recognition.  The ability to distinguish a tree
that has five apples growing on it from one that has two has an
obvious evolutionary advantage.

However, our main task is to explain why our most advanced and
abstract theories of mathematics crop up so often in modern physics,
not just the basic theories like arithmetic, and there is no
conceivable evolutionary pressure towards understanding the cosmos on
a large scale.  Knowing the ultimate fate of the universe may well be
crucial for our (very) long term survival, but cosmology operates on a
much longer timescale than evolution by natural selection, so, for
example, there is no immediate environmental pressure towards
discovering general relativity, nor the differential geometry needed to
formulate it.

On the other hand, it happens that evolution has endowed us with
general curiosity and intelligence.  This does have a survival
advantage as, for example, a species that is able to detect patterns
in predator attacks and pass that knowledge on to the next generation
without waiting for genetic changes to make the knowledge innate can
adapt to its environment more quickly.  Such curiosity and
intelligence are not the inevitable outcome of evolution, as the
previous dominance of dinosaurs on this planet demonstrates, but just
one possible adaptation that happened to occur in our case.  Like many
adaptations, it has side effects that are not immediately related to
our survival, one of which is that some of us like to think about the
large scale cosmos.

Since our general curiosity and intelligence are only a side effect of
adaptation, so are modern physics and mathematics.  Therefore, I find
it puzzling to argue for the efficacy of our reasoning in these areas
based on evolution.  Evolution often endows beliefs and behaviours
that are good heuristics for the cases commonly encountered, but a
poor reflection of reality (consider optical illusions for example).
With general intelligence one can, with considerable effort, reason
oneself out of such beliefs and behaviours.  This is why I stated that
mathematical intuition must be a product of general intelligence
rather than a direct evolutionary adaptation in my essay.

I think that understanding how a network of intelligent beings go
about organizing their knowledge is at the root of the efficacy of
mathematics in physics.  It should not matter whether those beings are
the products of evolution by natural selection or some hypothetical
artificial intelligences that we may develop in the future.  For this
reason, I take the existence of intelligent beings as a starting
point, rather than worrying about how they got that intelligence.

\subsubsection{Selection bias}

Wenmackers' selection bias argument is an attempt to deflate the idea
that mathematics is unreasonably effective in physics.  The idea is
that, due to selection bias, we tend to remember and focus on those
cases in which mathematics was successfully applied in physics,
whereas the vast majority of mathematics is actually completely
useless for science.  Here is the argument in her own words.

\begin{quotation}
  Among the books in mathematical libraries, many are filled with
  theories for which not a single real world application has been
  found.  We could measure the efficiency of mathematics for the
  natural science as follows: divide the number of pages that contain
  scientifically applicable results by the total number of pages
  produced in pure mathematics.  My conjecture is that, on this
  definition, the efficiency is very low.  In the previous section we
  saw that research, even in pure mathematics, is biased towards the
  themes of the natural sciences.  If we take this into account, the
  effectiveness of mathematics in the natural sciences does come out
  as unreasonable –- unreasonably low, that is.
\end{quotation}

My first response to this is to question whether the efficiency is
actually all that low.  After all, the vast majority of pages written
by theoretical physicists might also be irrelevant to reality, and
these are people who are deliberately trying to model reality.  We
need only consider the corpus of mechanical models of the ether from
the 1800's to render this plausible, let alone the vast array of
current speculative models of cosmology, particle physics, and quantum
gravity.  It is not obvious to me whether the proportion of applicable
published mathematics is so much lower than the proportion of correct
published physics, and, if it is not, then a raw page count does not
say much about the applicability of mathematics in particular.

Even if the efficiency of mathematics is much lower than that of
physics, it not obvious how low an inefficiency would be unreasonably
low.  If mathematics were produced by monkeys randomly hitting the
keys of typewriters then the probability of coming up with applicable
mathematics would be ridiculously small, akin to a Boltzmann brain
popping into existence via a fluctuation from thermal equilibrium.  In
light of such a ridiculously tiny probability, an efficiency of say
$0.01 \%$, which looks small from an everyday point of view, would
indicate an extremely high degree of unreasonable effectiveness.  Of
course, mathematicians are not typewriting monkeys, but unless one is
already convinced that the development of mathematics is correlated
with the development physics by one of Wenmackers' other arguments,
then even a relatively tiny efficiency could seem extremely
improbable.

My second response to the selection bias argument is that mathematics
is not identical to the corpus of mathematical literature laid out in
a row.  Some mathematical theories are considered more important than
others, e.g.\ the core topics taught in an undergraduate mathematics
degree.  Therefore, mathematical theories ought to be weighted with
their perceived importance when calculating the efficiency of
mathematics.  If you buy my network model of knowledge then the number
of inbound links to a node could be used to weight its importance, as
in the Google Page rank algorithm.  I would conjecture that the
efficiency of mathematics is much higher when weighted by perceived
importance.  I admit that this argument could be accused of
circularity because one of the reasons why an area of mathematics
might be regarded as important is its degree of applicability.
However, this just reinforces the point that mathematics not an
isolated subject, but must be considered in the context of the whole
network of human knowledge.

\subsection{Responses to comments on the FQXi website}

\subsubsection{Other processes in the knowledge network}

Several commenters expressed doubts that my theory of theory building
captures everything that is going on in mathematics.  For example,
Wenmackers commented:

\begin{quotation}
  Is this is correct summary of your main thesis (in section 4)? :
  "First, humans studied many aspects the world, gathering
  knowledge. At some point, it made sense to start studying the
  structure of that knowledge. (And further iterations.) This is
  called mathematics."

  Although I find this idea appealing (and I share your general
  preference for a naturalistic approach), it is not obvious to me
  that this captures all (or even the majority) of mathematical
  theories. In mathematics, we can take anything as a source of
  inspiration (including the world, our the structure of our knowledge
  thereof), but we are not restricted to studying it in that form: for
  instance, we may deliberately negate one of the properties in the
  structure that was the original inspiration, simply because we have
  a hunch that doing so may lead to interesting mathematics. Or do you
  see this differently?
\end{quotation}

There are other processes going on in the knowledge network beyond the
theory-building process that I described in my essay.  I did not
intend to suggest otherwise.  The reason why I focused on the process
of replacing direct links by more abstract theories is because I think
it can explain how mathematics becomes increasingly abstract, whilst
maintaining its applicability.  But this is clearly not the only thing
that mathematicians do.

One additional process that is going on is a certain amount of free
play and exploration, as also noted by Ken Wharton in his essay
\cite{Wharton2015}.  Mathematical axioms may be modified or negated to
see whether they lead to interesting theories.  However, as I argued
earlier, mathematical theories should be weighted with their perceived
importance when considering their place in the corpus of human
knowledge.  Modified theories that are highly connected to existing
theories, or to applications outside of mathematics, will ultimately
be regarded as more important.  It is possible that a group of pure
mathematicians will end up working on a relative backwater for an
academic generation or two, but this is likely to be given up if no
interesting connections are forthcoming.

For my theory, it is important that these additional processes should
not have a significant impact on the broad structure of the knowledge
network.  There should not be a process where large swaths of pure
mathematicians are led to work on completely isolated areas of the
network, developing a large number of internal links that raise the
perceived importance of their subject, with almost no links back to
the established corpus of knowledge.  Personally, I think that any
such process is likely to be dominated by processes that do link back
strongly to existing knowledge, but this is an empirical question
about how the mathematical knowledge network grows.  To address it, I
would need to develop concrete models, and compare them to the actual
growth of mathematics.

\subsubsection{What physical fact makes a mathematical fact true?}

Perhaps the highlight of the comment thread was a discussion with Tim
Maudlin.  It started with the following question:

\begin{quotation}
  I'm not sure I understand the sense in which mathematics is supposed
  to be ``about the physical world'' as you understand it. In one sense,
  the truth value of any claim about the physical world depends on how
  the physical world is, that is, it is physically contingent. Had the
  physical world been different, the truth value of the claim would be
  different. Now take a claim about the integers, such as Goldbach's
  conjecture. Do you mean to say that the truth or falsity of
  Goldbach's conjecture depends on the physical world: if the physical
  world is one way then it is true and if it is another way it is
  false? What feature of the physical world could the truth or falsity
  of the conjecture possibly depend on?
\end{quotation}

I stated in the essay that I think mathematical theories are formal
systems, but not all formal systems are mathematics.  The role of
physics is to help delineate which formal systems count as
mathematics.  Therefore, if Goldbach's conjecture is a theorem of
Peano arithmetic then I would say it is true in Peano arithmetic.   If
we want to ask whether it is true of the world then we have to ask if
Peano arithmetic is the most useful version of arithmetic by looking
how it fits into the knowledge network as a whole.  It may be that
more than one theory of arithmetic is equally useful, in which case we
may say that it has indefinite truth value, or we may want to say that
it is true in one context and not another if the two theories of
arithmetic have fairly disjoint domains of applicability.

If the Goldbach conjecture is not provable in Peano arithmetic, but is
provable in some meta-theory, then we can ask the same questions at
the level of the meta-theory, i.e.\ is it more useful than a different
meta-theory.  This sort of consideration has happened in mathematics
in a few cases, e.g.\ most mathematicians choose to work under the
assumption that the axiom of choice is true, mainly, I would argue,
because it leads to more useful theories.

At this point in the discussion, Maudlin accused me of being a
mathematical platonist.  His point is that if I accept theoremhood as
my criterion of truth then I am admitting some mathematical intuitions
as self-evident truths, so if my goal is to remove intuition as the
arbiter of mathematical truth then I have not yet succeeded.
Specifically, in order to even state what it means for something to be
a theorem in a formal system, I need to accept at least some of the
structure informal logic, e.g.\ things like: if A is true and B is
true then A AND B is also true.  Why do we accept these ideas of
informal logic?  Primarily because they seem to be self-evidently
true, but this is an appeal to unfettered mathematical intuition.

Maudlin points out that, because of this, it is difficult to avoid
some form of mathematical platonism, if only about the basic ideas of
logic.  I am not yet prepared to accept this and would argue, along
with Quine and Putnam \cite{Quine1951, Putnam1968}, that logic may be
empirical (in my case, I would say that even very basic informal logic
may be empirical).

I do not deny that I accept informal logic because it seems
self-evident to me.  However, as I argued in the essay, my intuitions
come from my brain, and my brain is a physical system.  Therefore, if
I have strong intuitions, they must have been put there by the natural
processes that led to the development of my brain.  The likely culprit
in this case is evolution by natural selection.  Organisms that
intuitively accept the laws of informal logic survive better than
those that do not because those laws are true of our physical
universe, so that creates a selection pressure to have them built in
as intuitions.

Does this mean there are conceivable universes in which the laws of
basic logic are different\footnote{The usual terminology for
  hypothetical universes with different laws is ``logically
  possible''.  That seems inappropriate here, but the terminology does
  show how ingrained into our minds the basic laws of logic are.},
e.g. in which Lewis Carroll's modus ponens denying tortoise
\cite{Carroll1895} is correct.  I have to admit that I have difficulty
imagining it, but it does not seem totally inconceivable.  In such a
universe, what counts as a mathematical truth would be different.  In
other words, mathematical truth might be empirical because the laws of
logic are.

In addition to this, there is a much more prosaic way in which
mathematical truth is dependent on the physical laws.  Imagine a
universe in which planets do not make circular motions as time
progresses, but instead travel in straight lines through a continually
changing landscape.  Inhabitants of such a planet would probably not
measure time using a cyclical system of minutes, hours, days, etc. as
we do, but instead just use a system of monotonically increasing
numbers.  A bit more fancifully, suppose that in this hypothetical
universe, every time a collection of twelve discrete objects like
sheep, rocks, apples, etc. are brought together in one place they
magically disappear into nothingness.  Inhabitants of this world would
use clock arithmetic, technically known as mod $12$ arithmetic, to
describe collections of discrete objects.  Their view of how to
measure time vs. how to measure collections of discrete objects would
be precisely the reverse of ours.  At least one of the senses in which
$12+1=13$ in our universe is not true in theirs, and I would say that
this is a sense in which mathematical truth depends on the laws of
physics.

It is fair to say that Maudlin was not impressed by this example, but
I take it deadly seriously.  What counts as mathematics and what
counts as mathematical truth are, in my view, pragmatically dependent
on how our mathematical theories fit into the structure of human
knowledge.  If the empirical facts change then so does the structure
of this network.  The meaning of numbers, in particular, is dependent
on how collections of discrete objects behave in our universe, and if
you change that then you change what makes a given theory of number
useful, and hence true in the pragmatic sense.  It is this that makes
the theory of number a hub in the network of human knowledge and this
is what philosophers ought to be studying if they want to understand
the meaning of mathematics.  The usual considerations in the
foundations of mathematics, such as deriving arithmetic from set
theory, though still well-connected to other areas of mathematics, are
comparative backwaters.  If we really want to understand what
mathematics is about, we ought to get our heads out of the formal
logic textbooks and look out at the physical world.

\subsubsection{Why are there regularities at all?}

Sophia Magnusdottir points out that my approach does not address why
there are regularities in nature to begin with.

\begin{quotation}
  In a nutshell what you seem to be saying is that one can try to
  understand knowledge discovery with a mathematical model as well. I
  agree that one can do this, though we can debate whether the one you
  propose is correct. But that doesn't explain why many of the
  observations that we have lend themselves to mathematical
  description. Why do we find ourselves in a universe that does have
  so many regularities? (And regularities within regularities?) That
  really is the puzzling aspect of the ``efficiency of mathematics in
  the natural science''. I don't see that you address it at all.
\end{quotation}

There are two relevant kinds of regularities here: the regularities
described by our most abstract mathematical theories on the one hand,
and the regularities of nature on the other.  On the face of it, these
two types of regularity have little to do with one another.  The fact
that the regularities described by our most abstract mathematical
theories so often show up in physics is what I take to be the
``unreasonable'' effectiveness of mathematics in the physical
sciences.

What I have tried to do is to argue that these two types of regularity
are more closely connected than we normally suppose.  They both
ultimately describe regularities, within regularities, \ldots, within
the natural world.  I have not even tried to address the question of
why there are regularities in nature in the first place.  Instead, I
have taken their existence as my starting point.  If we live in a
universe with regularities, the process of knowledge growth is such
that what we call mathematics will naturally show up in physics.  This
answers what I take to be the problem of ``unreasonable''
effectiveness.

Of course, one can try to go further by asking why there are any
regularities in the first place.  I do not think that anyone has
provided a compelling answer to this, and I suspect that it is one of
those questions that just leads to an infinite regress of further
``why'' questions.

For example, the Mathematical Universe Hypothesis may seem,
superficially, to explain the existence of regularities.  If our
universe literally is a mathematical structure, and mathematical
structures describe regularities, then there will necessarily be
regularities in nature.  However, one can then ask why our universe is
a mathematical structure, which is just the same question in a
different guise.

If we are to take the results of science seriously, the idea that our
universe is sufficiently regular to make science reliable has to be
assumed.  There is no proof of this, despite several centuries of
debate on the problem of induction.  Although this is an interesting
issue, I doubt that the problem can ever be resolved in an
uncontroversial way and it seems, to me at least, to be a different
and far more difficult problem than the ``unreasonable'' effectiveness
of mathematics in physics.  If my ideas are correct then at least
there is now only one type of regularity that needs to be explained.

\subsubsection{Elegance or efficiency?}

Alexy Burov made the following point.

\begin{quotation}
  Wigner's wonder about the relation of physics and mathematics is not
  just abut the fact that there are some mathematical forms describing
  laws of nature. He is fascinated by something more: that these forms
  are both elegant, while covering a wide range of parameters, and
  extremely precise. I do not see anything in your paper which relates
  to that amazing and highly important fact about the relation of
  physics and mathematics.
\end{quotation}

I take the key issue here to be that I have not explained why the
mathematics used in physics is ``elegant''.  After all, if we had a
bunch of different laws covering different parameter ranges then we
could always put them together into a single structure by inserting a
lot of ``if'' clauses into our laws of physics.  We can also make them
arbitrarily precise by adding lots of special cases in this way.
Presumably though, the result of this would be judged ``inelegant''.

To be honest, I have a great deal of trouble understanding what
mathematicians and physicists mean by ``elegance'' (hence the scare
quotes).  For this reason, I have emphasized that the mathematics in
modern physics is ``abstract'' and ``advanced'' rather than
``elegant''.

A more precise definition of elegance is needed to make any progress
on this issue.  One concrete suggestions is that perhaps elegance
refers to the fact that the fundamental laws of physics are few in
number so they can be written on a t-shirt.  It is tempting to draw
the analogy with algorithmic information here, i.e.\ the length of the
shortest computer program that will generate a given output
\cite{Grunwald2008}.  Perhaps the laws of physics are viewed as
elegant because they have low algorithmic information.  We get out of
them far more than we put in.

So, perhaps what we call ``elegance'' really means the smallest
possible set of laws that encapsulates the largest number of
phenomena.  If so, then what we need to explain is why the process of
scientific discovery would tend to produce laws with low algorithmic
information.  The idea that scientists are trying to optimize
algorithmic information directly is a logician's parody of a complex
social process.  Instead, we need to determine whether the processes
going on in the knowledge network would tend to reduce the algorithmic
information content of the largest hubs in the network.  In this I am
encouraged by the fact that many scale-free networks exhibit the
``small world'' phenomenon in which the number of links in a path
connecting two randomly chosen nodes is small\footnote{Technically, it
grows like the logarithm of the number of nodes.}.  If this is true of
the knowledge network then it means that the hubs must be powerful
enough to derive the empirical phenomena in a relatively small number
of steps.  The average path length between two nodes might be taken as
a measure of the efficiency with which our knowledge is encoded in the
network, or, if you prefer, its ``elegance''.

Now, of course, this may be completely unrelated to what everyone else
means by the word ``elegance'', as applied to mathematics and physics.
If so, a more precise definition, or an analysis into more primitive
concepts, is needed before we can address the problem.  Once we have
that, I suspect the problem might not look so intractable.

\bibliographystyle{unsrturl}
\bibliography{FQXi2015}

\end{document}